# Large Deflections of Beam Subject to Three-points Bending

Milan Batista
University of Ljubljana, Faculty of Maritime Studies and Transport
Pot pomorščakov 4, 6320 Portorož, Slovenia
milan.batista@fpp.uni-lj.si

**Abstract**

In the paper a solution for equilibrium configurations of an elastic beam subject to three points bending is given in terms of Jacobi elliptical functions. General equations are derived and the domain of solution is established. Several examples that illustrate a use of the solution are discussed. The obtained numerical results are compared with results of other authors. Approximation formula by which the beam load is given as polynomial function of beam deflection is also derived. The range of applicability of the approximation is illustrated by numerical example.

## 1 Introduction

In this article we will treat the problem of determination of the relation between the force and deflection for the three-point symmetric bending of a thin elastic beam. The problem is well-known and is closely related to the three points bending test by which one can indirectly measure flexural properties of beam material [1].

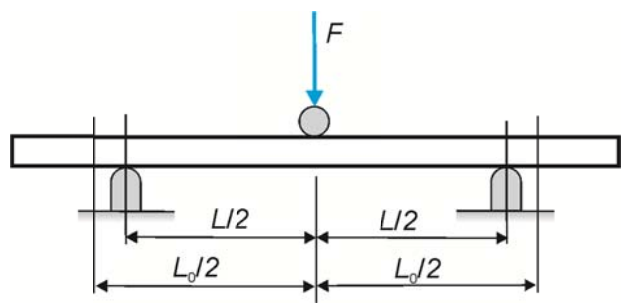

**Figure 1.** Three points bending geometry and load

The solution for small deflected beam is well known and is given by

$$\frac{\delta}{L} = \frac{FL^2}{48EI} \tag{1}$$

where $\delta$ is the midpoint deflection, $L$ is span between supports, $F$ is applied force, $E$ is Young's modulus and $I$ is the area moment of inertia of the beam cross-section. For large deflections among 14/08/2014 09:12arXiv:1408.3094 [physics.class-ph]first the problem was treated by Freeman [2]. In the paper he from assumption that bending moment is proportional to curvature of beam central surface derives two nonlinear equations which relate applied force, span between supports and midpoint deflection as function of beam tangent angle at supports. He solves these equations graphically for various combinations of given and unknown parameters that configure in these equations. He discussed possible forms of beam with constant span as function of tangent angle at supports and also gives some experimental verification of his theoretical results. Similar relations, but by using elliptic integrals, was obtained by Conway [3] who also study the effect of friction at the supports to the beam midpoint deflection. In particular he for given tangent angle at support point calculates corresponding load and midpoint deflection and present result in graphical form. An analysis of effect of friction can be found also in Frisch-Fay book [4] where the author shows how the solution of the problem can be obtained from the solution for cantilever beam. The problem in context of applying standardization of three point bending test for large elastic deflection was treated by West [5]. He numerically solves nonlinear beam equation and present midpoint deflection in the form of correction factors to linear beam theory for various values of load. He also observed that finite diameter of the roller supports change the effective span, however he use this fact only to correct his experimental data. We note that previous geometry study of span shortening was done by Westwater [6]. The effect of friction and radius of the supports on the solution of beam equations was treated by Theocaris et al. [7]. The authors also study the effect of thickness $h$ of beam on the position of the beam neutral axis. They show that due to an axial force neutral axis is not the beam central axis as was for example explicitly assumed by Freeman [2] but rather a new unknown of the problem. By numerical calculations they show that effect of neutral axis displacement is negligible for beams with ratio say $h/L \geq 20$. We note that the authors do not include effect of supports radius into the calculations but rather they give some correction formulas. Among authors that discuss the problem in last century we mention Ohtsuki [8] who among other perform some comparisons of results of beam bending obtained by analytical, numerical and experimental method, and found well agreement. We note that he does not include friction nether finite suports diameter in his calculations.

Now, in typical three point test the measured quantities are force and deflection from which one can according to some underlying theory calculates various material characteristics. While authors from the previous century for these calculations offers a corrections to the equations by a numerical factors or in a graphical form, the appearance of low cost computers at 1990s make such a methods obsolete. Thus Arnautov [9] give a solution of the problem in the form of elliptic integrals and conduct the calculation of various data from the measured force and deflection by numerical procedure. He however neglect support radius and possible friction. Also, apart not dealing with large deflections, we mention Mujika paper [10] where he study the support span reduction due to rotation at supports on calculation of bending modulus. We note that he assume that rotation angle is small and thus he neglected vertical displacement of contact points.

At the end of the review we add that it is clearly incomplete since there is a huge amount of articles treating various aspects of three points bending of beams but we want to address only those which directly influence this paper. In particular we didn't include papers which treat simply supported beams and papers where the beam is loaded by distributed load.

From the review we see that the problem of three-points beam bending has been well studied but it seems that there are still some questions not yet answered. For example none of the mentioned authors discusses the domain of the solution neither has authors try to derivation an approximate formulas from the general solution. Thus in this article the problem will be discussed once more. In what follows we will first give a general solution of the problem in terms of Jacobi elliptical function. Like Coleman [3], Frisch-Fay [4] and Theocaris et al. [7] we will treated both the smooth and the rough supports but unlike mention authors we will also discuss effect of radius of the supports in full details. Using general solution we will then establish its domain and give few numerical examples. At the end of the article we will use the general solution to derive an approximate formula which give explicate relation between the beam load and its midpoint deflection. For all calculations that follow we use the Maple computer algebra program.

## 2 Formulation of the problem and its solution

We consider initially straight inextensible elastic beam subject to vertical force acting at the middle of the supports (see Figure 1). The force produces vertical deflection of the beam and also move contact points towards midpoint [2]. From the geometry of deformed beam shown on Figure 2 we can derive that on a round support each of the contact point move in horizontal and vertical direction respectively by

$$\Delta_x = r\sin\phi_0 \quad \text{and} \quad \Delta_y = r(1-\cos\phi_0) \qquad (2)$$

where $r$ is the radius of supports and $\phi_0$ is the beam tangent angle at supports.

From the equilibrium of beam in vertical direction we found that reaction force $R$ at each support is given by

$$R = \frac{F}{2\sin\alpha} \qquad (3)$$

where $\alpha$ is unknown reaction force angle (Figure 2). The reaction $R$ can be with respect to beam base curve always resolved to the normal component $N$ and the tangential component (friction force) $T$ so that

$$|R| = \sqrt{N^2 + T^2} \quad \text{and} \quad \frac{T}{N} = \tan\gamma = \mu \qquad (4)$$

where $\mu$ is coefficient of stick. Here we must distinguish three cases:

- If supports are smooth then $T = 0$ and therefore $\mu = 0$. The only unknowns in this case are $\alpha$ and $N$.
- If supports are perfectly rough then we have three unknowns are $\alpha$, $N$ and $T$ (or $\mu$)
- If supports are rough then we must impose some friction law. In particular by Coulomb friction law the equilibrium is maintained for $\mu \leq \mu_s$ where $\mu_s$ is coefficient of static friction. When $\mu < \mu_s$ we again have three unknowns: $\alpha$, $N$ and $T$. When this value is exceed the beam slides over support rollers and $\mu$ became coefficient of kinetic friction $\mu_k$ which has approximately constant value.

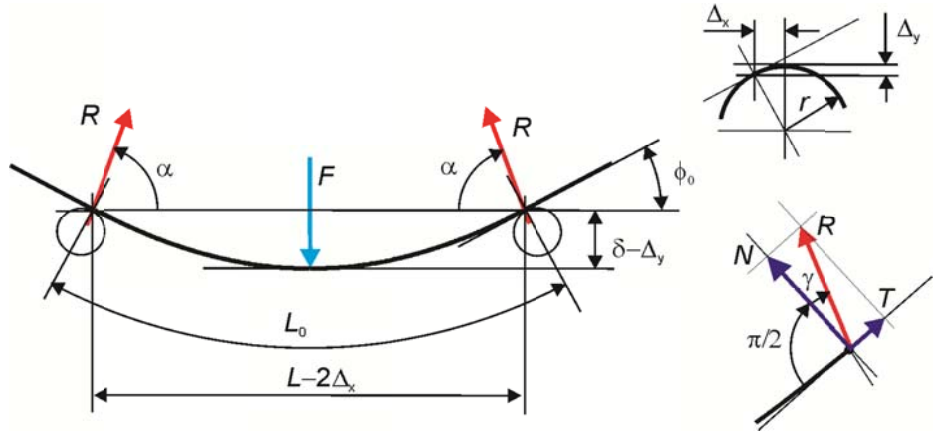

**Figure 2.** Equilibrium of the deformed beam

We note, as was already observed by Conway [3], that for even more sophisticated analysis we should include into consideration also equivalent bending moment at the beam contact points. Namely if we translate reaction force $R$ from contact point to the point on base curve then we must there also add the bending moment $M = hT/2 = \mu hN/2$ where $h$ is beam thickness. However we assume that $h$ is in some sense small so this moment is assumed to be negligible. For more details on this effect see Theocaris at all paper [7].

Now the discussed problem is because of its symmetry equivalent to the problem of cantilever subject to the force $R$ [4]. The equations describing the deformed cantilever base curve are known and are given by the following parametric equations [11]

$$\frac{x(s)}{\ell_0} = \xi(s)\cos\alpha + \eta(s)\sin\alpha \qquad \frac{y(s)}{\ell_0} = -\xi(s)\sin\alpha + \eta(s)\cos\alpha \qquad (0 \le s \le 1) \quad (5)$$

where

$$\xi(s) = \left[\frac{2\mathrm{E}(k)}{\mathrm{K}(k)} - 1\right](1-s) + \frac{2}{\omega}\left[Z(\omega + \mathrm{K}, k) - Z(\omega s + \mathrm{K}, k)\right] \quad (6)$$

$$\eta(s) = -\frac{2k}{\omega}\left[\mathrm{cn}(\omega + \mathrm{K}, k) - \mathrm{cn}(\omega s + \mathrm{K}, k)\right]$$

$$\sin\frac{\alpha}{2} = k\,\mathrm{sn}(\omega + \mathrm{K}, k) \quad (7)$$

$$k \equiv \sin\frac{\theta_0}{2} \quad (8)$$

$$\omega^2 \equiv \frac{R\ell_0^2}{EI} \quad (9)$$

Here $s$ is normalized arc length parameter, $\mathrm{K} = \mathrm{K}(k)$ and $\mathrm{E} = \mathrm{E}(k)$ are respectively complete elliptic integrals of first and second kind, $sn$, $cn$ and $Z$ are respectively Jacobi sine, cosine and zeta elliptic function, *EI* is the bending stiffness of the cantilever and $\ell_0 \equiv L_0/2$ is its length. The geometric meaning of $\xi(s)$ and $\eta(s)$ and the angels $\theta_0$ should be clear from Figure 3. We note that effective procedures exist for calculation of Jacobi's elliptic functions [12, 13] and also that they are available in a commercial programs such as Matlab and Maple.

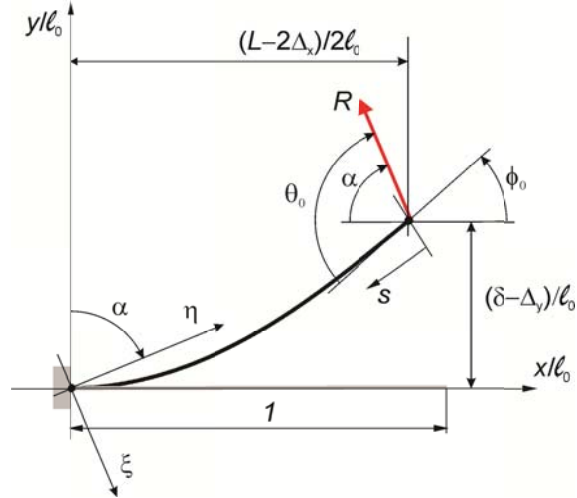

**Figure 3.** Geometry and load of the cantilever problem

Alternatively we can by using Eq (7) and trigonometric identities express $\sin\alpha$ and $\cos\alpha$ which appears in Eqs (5) as follows

$$\sin\alpha = 2\sin\frac{\alpha}{2}\sqrt{1-\sin^2\frac{\alpha}{2}} = 2k\,\mathrm{sn}(\omega+\mathrm{K},k)\,\mathrm{dn}(\omega+\mathrm{K},k)$$

(10)

$$\cos\alpha = 1 - 2\sin^2\frac{\alpha}{2} = 1 - 2k^2\,\mathrm{sn}^2(\omega+\mathrm{K},k)$$

In this way we see that the shape of deformed beam given by Eqs (5) is completely determinate once parameters $\omega$ and *k* are known. The next task is thus to connect these parameters with the deformed beam geometry and the beam load.

Now, to calculate change in support span $\Delta_x$ and change in deflection $\Delta_y$ which are given by Eq (2) we need to know the free end tangent angle $\phi_0$. Using Eq (7) and Eq (8) we find

$$\phi_0 = \theta_0 - \alpha = 2\left\{\sin^{-1}k - \sin^{-1}\left[k\,\mathrm{sn}(\omega+\mathrm{K},k)\right]\right\}$$

(11)

Next, from Eqs (5) we by setting $s=0$ obtain the non-dimensional coordinates of cantilever free end

$$\bar{x}_0 \equiv \frac{x(0)}{\ell_0} = \xi_0\cos\alpha + \eta_0\sin\alpha \qquad \bar{y}_0 \equiv \frac{y(0)}{\ell_0} = -\xi_0\sin\alpha + \eta_0\cos\alpha$$

(12)

where by Eqs (6)

$$\xi_0 \equiv \xi(0) = \frac{2E}{K} - 1 + \frac{2}{\omega} Z(\omega + K) \qquad \eta_0 \equiv \eta(0) = -\frac{2k}{\omega} \text{cn}(\omega + K) \qquad (13)$$

In order to normalize the beam coordinates to known span *L* we must first obtain expression for the unknown half beam length $\ell_0$. By using definition $\bar{x}_0 = \frac{x(0)}{\ell_0}$ and geometry relation $x(0) = (L - 2\Delta_x)/2$ we find

$$\frac{\ell_0}{L} = \frac{1 - 2\Delta_x/L}{2\bar{x}_0} \qquad (14)$$

Similarly, to express deflection $\delta$ normalized to *L* rather than to $\ell_0$ we use definition $\bar{y}_0 = \frac{y(0)}{\ell_0}$ and geometric relation $y(0) = \delta - \Delta_y$ so

$$\frac{\delta}{L} = \bar{y}_0 \frac{\ell_0}{L} + \frac{\Delta_y}{L} \qquad (15)$$

Further, definition (9) and equilibrium condition (3) give

$$\frac{FL^2}{EI} = \frac{2\omega^2 \sin\alpha}{(\ell_0/L)^2} \qquad (16)$$

Finally, from geometry (Figure 2 and 3) we have

$$\theta_0 = \frac{\pi}{2} + \gamma \qquad (17)$$

and therefore by Eqs (4) and (8) we obtain

$$k = \sin\left(\frac{\pi}{4} + \frac{\tan^{-1}\mu}{2}\right) = \frac{\sqrt{2}}{2}\sqrt{1 + \frac{\mu}{\sqrt{1+\mu^2}}} \qquad (18)$$

Alternatively by inverting this expression we find

$$\mu = \frac{2k^2 - 1}{2k\sqrt{1-k^2}} \qquad (19)$$

In this way the variables $\alpha$, $\phi_0$, $\ell_0/L$, $\delta/L$, $FL^2/EI$ and also $\Delta_x$ and $\Delta_y$ which describe the deformed beam are given respectively by expressions (7), (11), (14), (16) and (2) where $\omega$ and $\mu$ plays role of parameters. Once the values of those parameters are chosen the values of the variables can be calculated in straightforward way. We also note that all the variables except $\alpha$ and $\phi_0$ depend also on supports radius *r/L*.

**Note.** When $\omega$, $\mu$, *r* and *L* are given the space coordinates of deformed beam are calculate by

$$x(s) = \ell_0 \left[ \xi(s)\cos\alpha + \eta(s)\sin\alpha \right]$$

$$y(s) = \ell_0 \left[ -\xi(s)\sin\alpha + \eta(s)\cos\alpha \right] - \delta \quad (0 \le s \le 1) \tag{20}$$

wherein the centre of supports are at

$$x = \pm L/2 \quad y = -r \tag{21}$$

### 3 The domain of the solution

The domain of the solution must be derived from the assumption of the problem that $F \ge 0$ and also that unilateral contact between the beam and supports require that $R \ge 0$ (Figure 2). The last inequality by equilibriun (3) imply condition $0 \le \alpha \le \pi$.

Now, by (16) we have $F = 0$ in the three cases:

- when $\omega = 0$,
- when $\omega > 0$ and $\alpha = 0$ or
- when $\omega > 0$ and $\ell_0 = \infty$.

*Case $\omega = 0$.* In this case Eq (9) imply $R = 0$ and further by equilibrium condition (3) this can be when $F = 0$. This is the case of free beam.

*Case $\omega > 0$ and $\alpha = 0$.* In this case we from Eq (7) obtain $\text{sn}(\omega + \text{K}, k) = 0$ and from this we conclude that $\omega \le \text{K}(k)$. For the limit case when $\omega = \text{K}(k)$ we have $F = 0$. This means that any further increase of force cause that already deformed beam slip through between supports. Also for the case $\omega = \text{K}(k)$ we from Eqs (14) and (15) obtain expressions for required (maximum) beam length and its maximum possible midpoint deflection

$$\frac{L_{0\max}}{L} = \frac{2\ell_0}{L} = \frac{\text{K}}{2\text{E}-\text{K}}\left(1 - 4\frac{r}{L}k\sqrt{1-k^2}\right) \tag{22}$$

$$\frac{\delta_{\max}}{L} = \frac{k}{2\text{E}-\text{K}}\left[1 + 2\frac{r}{L}k\left(2\sqrt{1-k^2} + K - 2\text{E}\right)\right] \tag{23}$$

It is seen from these relations that presence of support radius $r$ reduces $L_{0\max}$ and increase $\delta_{\max}$. The domain of $k$ is by definition (8) restricted to $0 \le k \le 1$. However the problem parameters impose additional restriction to $k$. First, since $\mu \ge 0$ we by Eq (18) have $k \ge \sqrt{2}/2$. Further, since $L_{0\max} > 0$ we for $r/L \le 1/2$ must have $2\text{E}(k) - \text{K}(k) > 0$. By Eq (23) this condition also ensures that $\delta_{\max} > 0$. Geometrically this means that deformed beam cannot form a closed loop [11]. The applicability of (22) and (23) is therefore bounded to the interval

$$\sqrt{2}/2 \le k < k_* \tag{24}$$

where $k_*$ is solution of $2E(k)-K(k)=0$ which is $k_* \approx 0.9089$. When $k$ approaching this upper limit both $L_{0max}$ and $\delta_{max}$ increase without bound. For $k_*$ we have corresponding $\omega_* \approx 174.1$ and $\mu_* \approx 0.8604$.

*Case* $\omega > 0$ *and* $\ell_0 = \infty$. By Eq (14) we in this case must have $\bar{x}_0 = 0$. The first of Eqs (12) then represent as highly nonlinear equation for unknown $\omega(k)$ which can be solved only numerically. From the graph on Figure 4 we see that the case is possible only for $k > k_*$ that is for $\mu \geq \mu_*$. We note that condition $\ell_0 = \infty$ means that the beam theoretically inflates without bounds and at the limit form a loop. However physically this means that the limit case $F = 0$ cannot be reached, that is, the beam slip through between supports because of its finite length rather because of applied force. The special case is when $k = 1$ which by Eq (19) give $\mu = \infty$. This is the case of perfectly rough supports where the initially straight beam remains straight under any applied force since the tangential component of reaction force prevents any movement of contact points.

On graph on Figure 4 the domain of solution is present as shaded region. Also on graphs on Figure 5 we present load given by (16) and deflection given by (15) as function of $\omega$ and $k$ for the case $r = 0$.

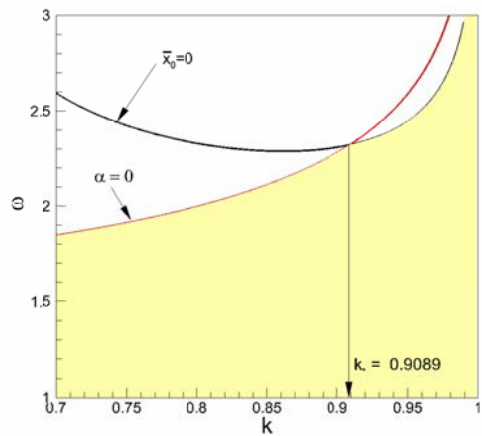

**Figure 4.** The domain of solution imposed by $F \geq 0$ is represented as shaded region on $(k, \omega)$ plane. On curves $\alpha = 0$ and $\bar{x}_0 = 0$ we have $F = 0$. Note that the domain is independent on supports radius.

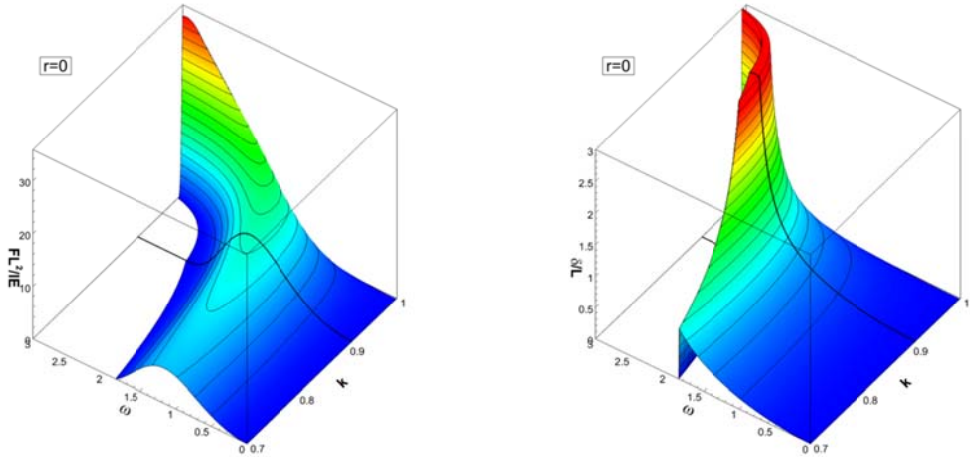

**Figure 5**. Beam load given by (16) and its deflection given by (15) as function of $\omega$ and $k$ when $r=0$.

### 4 Examples

On the base of the formulas derived in second section we can by choosing various problem variables as unknowns design various problems. However we will focus only to problems involving the load and the deflection of a beam. More specific, we can from Eqs (15) and (16), and the formulas given in the second section, deduce the following explicate relations

$$\frac{\delta}{L}=f_1(\omega,\mu,r/L) \quad \text{and} \quad \frac{FL^2}{EI}=f_2(\omega,\mu,r/L) \tag{25}$$

where we assume that span length $L$ and supports radius $r$ are given and where $f_1$ and $f_2$ are some functions of complex structure. These relations allow us to define for the following problems:

- When $\omega$ and $\mu$ are given then the unknowns are $\delta/L$ and $FL^2/EI$ which can be calculated by Eqs (25). This is direct problem.
- When $\delta/L$ (or $FL^2/EI$) and $\mu$ are given, then we have half inverse problem where the unknowns are $\omega$ and $FL^2/EI$ (or $\delta/L$). For given $\delta/L$ the first of Eqs (25) is nonlinear equation for unknown $\omega$. As can be seen from graph on Figure 6 deflection $\delta/L$ is monotone function of $\omega$ so this equation if it has solution then it is unique. Once $\omega$ is known we can calculate $FL^2/EI$ from the second of Eqs. (25). In a case of measurement $F$ is also given so we can future calculate beam stiffness $IE$. If $FL^2/EI$ is given then the second of Eqs (25) is nonlinear equation for unknown $\omega$. Bell shaped relations between $FL^2/EI$ and $\omega$ which can be observed on graph on Figure 5 and Figure 6 indicate that there are in this case up to two possible solutions of the equation.
- When $FL^2/EI$ and $\delta/L$ are given then relations (25) forms a system of two nonlinear equations for unknowns are $\omega$ and $\mu$. This is inverse problem. Again from the graphs on Figure 5 we conclude that if the system has solution then it is unique.

In the connection with these problems it is of order to clarify the role of stick coefficient $\mu$. We have three possibilities:

- If we consider sequences of beam deformations beginning from its initial state by steady increasing/decreasing of the force which produces rolling/sliding of the beam over support rollers then $\mu = \mu_k$
- For the case $\mu < \mu_s$ or for the case when supports are perfectly rough the problem is indeterminate in a sense that we cannot calculate deflection $\delta$ for given load $FL^2/EI$ because we don't have an equation for calculation of $\mu$. In other words the problem becomes indirect problem where the reaction forces are determinate by the applied force and by the geometry of deformed beam.
- The case $\mu = \mu_s$ is intermediate state where future increase of applied force cause the beam to slip between supports [4]

The rest of this section is devoted to some examples.

*Direct problem.* For the purpose of illustration of the relations derived in the second section the graphs of $\delta/L$ as function of $\omega$ and $\phi_0$, $\ell_0/L$ and $FL^2/EI$ as function of $\delta/L$ are for the case $\mu = 0$ and $r = 0$ displayed on Figure 6. These graphs shows that with increasing $\delta/L$ except for except $FL^2/EI$ we have monotonically increasing $\omega$, $\ell_0/L$ and $\phi_0$. The relation between $\delta/L$ and $FL^2/EI$ is well known bell shaped load curve with local maximum between two inflection points meaning that two deflections values are possible for one value of force. In Table 1 we also give some reference values which are of use when one develops his/her own program.

The present value of maximum normalized load and correspondent normalized deflection given in Table 1 agree with one given by Frisch-Fay [4] pp 75. To compare his values with present we replace his *P* by *F*/2 and his *L* by present *L*/2, that is, his values for load parameter must be multiplied by 8 and his values for deflection divided by 2. We thus have $FL^2/EI = 0.835 \times 8 = 6.680$ and $\delta/L = 0.4764/2 = 0.2382$ which agree with values in Table 1 for the point 2. The deformed beams shapes correspond to values from Table 1 are shown on Figure 7.

Future verification of the present method is given on Figure 8 where the load curve together with experimental points provided by West [5] is shown. His data set was taken because it is one of rear data set public available in tabular form. We note that it is not quite clear how West obtained his values so we take into account his note that $FL^2/EI$ was corrected by 20%. With this correction the discrepancy between calculated and measured values is for all point except last three less than 4%.

**Table 1.** Calculated values when $\mu = 0$ and $r = 0$

| Point | | $\omega$ | $\delta/L$ | $L_0/L$ | $FL^2/EI$ | $\phi_0$ | $\alpha$ |
|---|---|---|---|---|---|---|---|
| 1 | inflection | 0.73115 | 0.08980 | 1.01922 | 3.97140 | $15^0 16' 42''$ | $74^0 43'$ |
| 2 | maximum | 1.16505 | 0.23819 | 1.13015 | 6.67181 | $38^0 18' 04''$ | $51^0 42'$ |
| 3 | inflection | 1.60219 | 0.52019 | 1.54559 | 2.96785 | $69^0 48' 14''$ | $20^0 12'$ |
| 4 | zero | 1.85407 | 0.83467 | 2.18844 | 0 | $90^0$ | 0 |

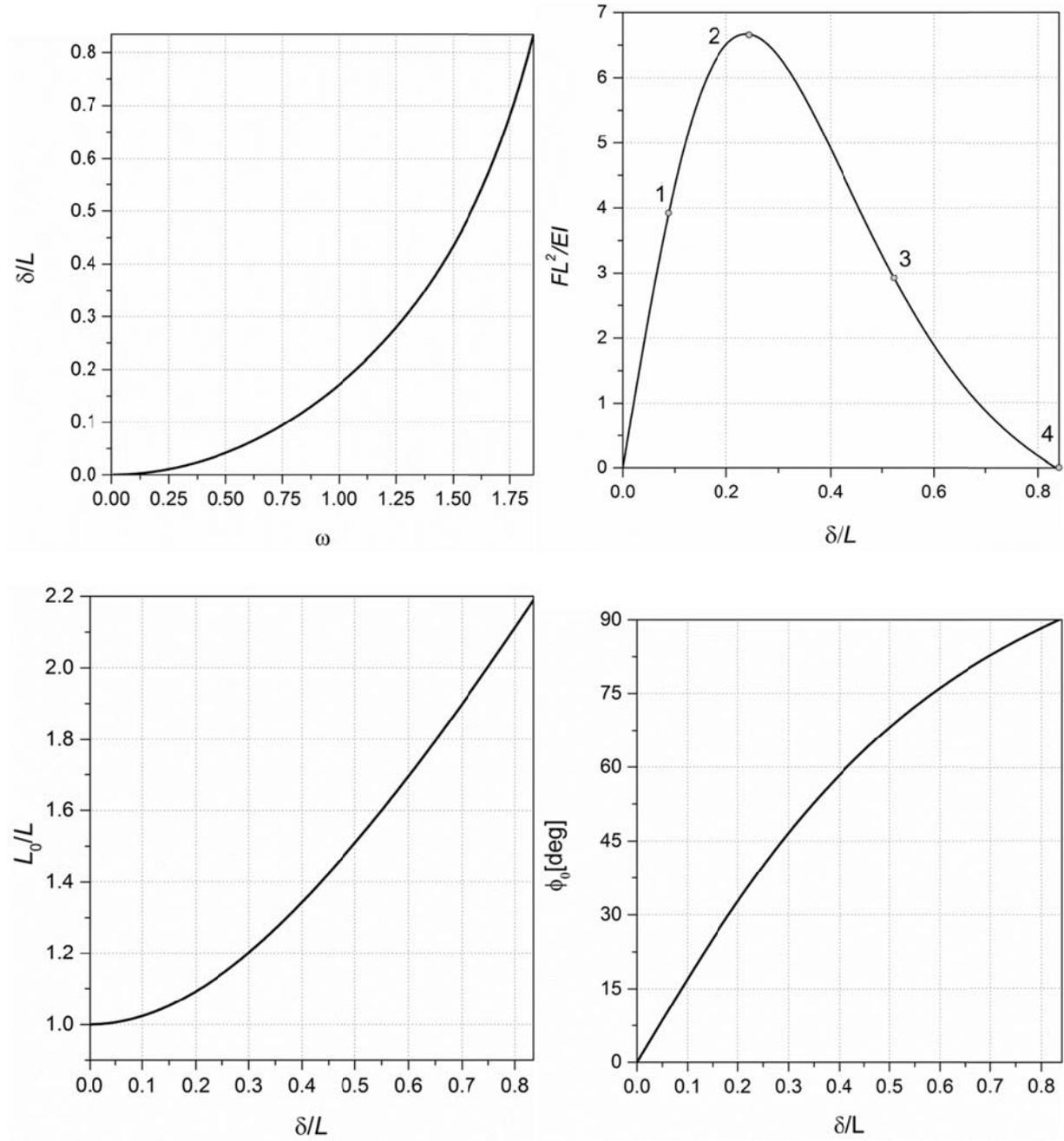

**Figure 6.** Graphs of dependences among various variables when $\mu = 0$ and $r = 0$

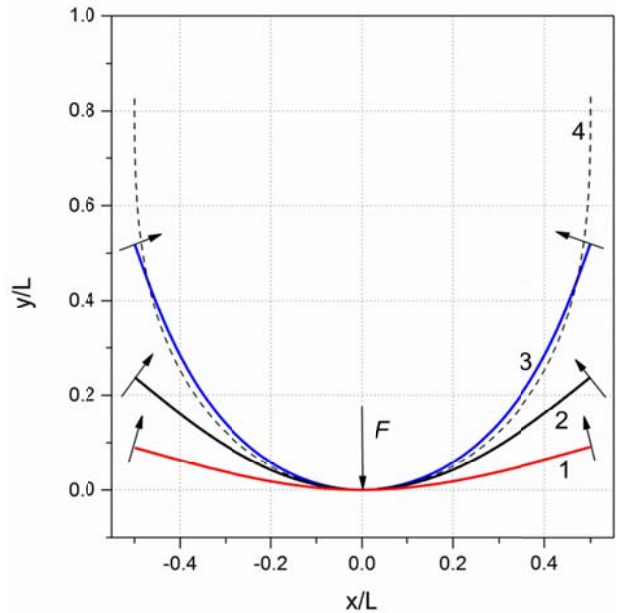

**Figure 7.** Deformed beam shapes for load cases from Table 1

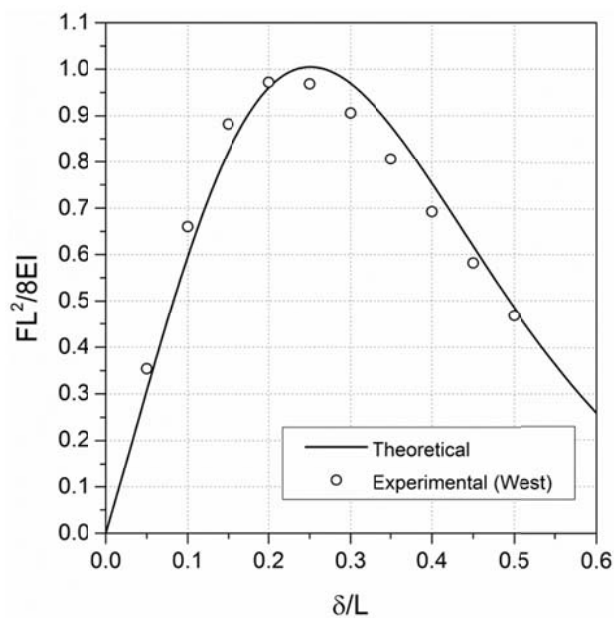

**Figure 8.** Load-deflection curve for smooth supports with $r/L = 0.07$.

*Half inverse problem.* Consider now typical calculation from experimental data. We assume that *L*, *r* and $\mu$ are given and *F* and $\delta$ are measured. Then we calculate the unknowns as follows:

1. Calculate $\delta/L$, $r/L$
2. Solve Eq (15) for unknown $\omega$ where $\omega = 1$ is good initial guess
3. Calculate bending stiffness *EI* from (16)
4. Calculate maximum moment at midpoint from [11]

$$M_{max} = EI \frac{\kappa(1)}{\ell_0} = -\frac{2EI\omega k}{L(\ell_0/L)} \text{cn}(\omega + K, k) \qquad (26)$$

5. Calculate bending stress in outer fibres at midpoint $\sigma_{max} = \dfrac{M_{max}}{W}$ where $W$ is the section modulus

The results of calculation for example from Arnutov [9] are given in Table 2. It is seen from the table a good agreement between his results and results obtain by present method. We note that Arnutov did not include the support radius in his calculations so radiuses given in Table 2 are hypothetical.

**Table 2.** Results of calculation for the example given by Arnutov [9]. Length of span between supports $L=30\,mm$, thickness of specimen $h=0.5\,mm$, width of specimen $b=6.57\,mm$, measured deflection at failure $\delta = 7.23\,mm$ and the corresponded measured force $F = 7.7\,dN$. $(\mu = 0)$

|  | $r$ mm | $\alpha$ | $\phi_0$ | $\ell_0$ mm | $R$ dN | $EI$ dNmm$^2$ | $E$ dN/mm$^2$ | $M_{max}$ dN mm | $\sigma$ MPa |
|---|---|---|---|---|---|---|---|---|---|
| Arnutov | 0 | $51.25^0$ | - | 17.001 | 4.973 | - | - | 80.125 | 2869.4 |
| Present | 0 | $51.30^0$ | $38.70^0$ | 16.997 | 4.933 | 1038.82 | 15184.4 | 80.048 | 2866.6 |
| Present | 0.5 | $51.11^0$ | $38.89^0$ | 16.662 | 4.946 | 995.94 | 14557.7 | 78.650 | 2816.6 |
| Present | 1.0 | $50.91^0$ | $39.09^0$ | 16.325 | 4.960 | 953.66 | 13939.7 | 77.234 | 2765.9 |
| Present | 1.5 | $50.71^0$ | $39.29^0$ | 15.984 | 4.974 | 912.00 | 13330.6 | 75.799 | 2714.5 |

*Inverse problem.* On Figure 9 a simple beam deflection experiment together with calculated shapes is shown. Despite the fact that the experiment was in some sense sloppy the agreement between realized and calculated beam shape is excellent. In experiment we us carbon fibre beam with while for supports we use two steel covers. By known load and measured deflection we by Eqs (25) calculate $\mu$ and $\omega$ (see Figure 10). In principle this is relatively easy task however, as it is seen from Figure 10, the intersecting curves for small deflection form relatively small angle at intersection point (left graph) or equivalently, possible range of solution is very narrow (right graph). Therefore the calculation of intersection become very sensitive on values of $FL^2/EI$ and $\delta/L$.

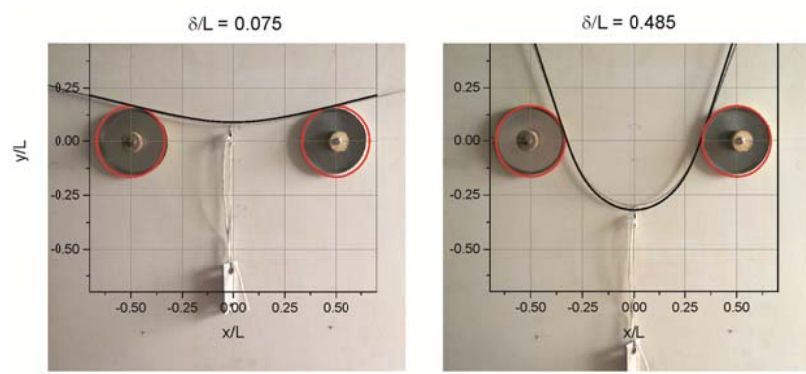

**Figure 9**. Overlapping of calculated (dark line) and realized carbon fibre beam shape (brighter line) for the case $r/L = 0.163$

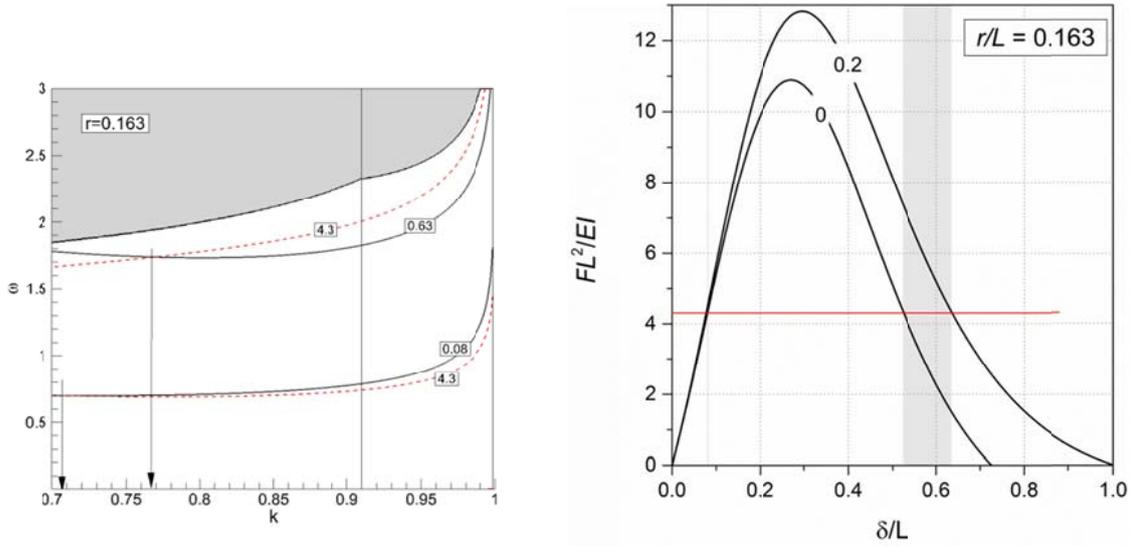

**Figure 10.** Calculation of the stick coefficient (left graph). Two cases when $FL^2/EI = 4.309$ (dashed line) are shown. For $\delta/L = 0.08$ we have $k = 0.71$ ($\mu \approx 0$) and for $\delta/L = 0.63$ we have $k = 0.768$ ($\mu = 0.184$). On the right graph the range of possible solutions for given $FL^2/EI$ is shown for $0 \leq \mu \leq 0.2$.

**Some approximation formulas**

The present solution does not allow to explicitly express $FL^2/EI$ as function of $\delta/L$ or otherwise. However these variables can be expanded into the power series of $\omega$ and these series can be then inverted [14]. The procedure for obtaining the desired explicit relation is thus the following.

First we by using Eqs (15) and (16) expand

$$\frac{FL^2}{EI} = \sum_{n=1}^{N} a_n(\mu, r) \omega^{2n} + \cdots \quad \text{and} \quad \frac{\delta}{L} = \sum_{n=1}^{N} b_n(\mu, r) \omega^{2n} + \cdots \quad (27)$$

Now, since the relation between $\delta/L$ and $\omega$ is monotone, the inversion of second series is more suitable. Thus we set

$$\omega^2 = \sum_{m=1}^{M} c_m \left(\frac{\delta}{L}\right)^m + \cdots \quad (28)$$

where $c_m$ are unknown coefficients. Substituting this into second of series (27) we obtain

$$\frac{\delta}{L} = \sum_{n=1}^{N} b_n \left( \sum_{m=1}^{M} c_m \left(\frac{\delta}{L}\right)^m \right)^{2n} + \cdots \quad (29)$$

Comparing the coefficients at equal powers of $\frac{\delta}{L}$ yield recursive infinite system of linear algebraic equations for unknown $c_m$. Once $c_m$ are known we by substituting series (28) into first of series (27) obtain

$$\frac{FL^2}{EI} = \sum_{n=1}^{N} a_n \left( \sum_{m=1}^{M} c_m \left(\frac{\delta}{L}\right)^m \right)^{2n} + \cdots \quad (30)$$

For practical calculation we use symbolic manipulation program Maple. The result of calculation is the following series expansion

$$\frac{FL^2}{48EI} = \left(\frac{\delta}{L}\right) + \left(\frac{12}{5}\mu + \frac{27}{2}\frac{r}{L}\right)\left(\frac{\delta}{L}\right)^2 - \left(\frac{72}{7} + \cdots\right)\left(\frac{\delta}{L}\right)^3 - \left(18\mu + \frac{11475}{56}\frac{r}{L} + \cdots\right)\left(\frac{\delta}{L}\right)^4$$
$$+ \left(\frac{37260}{539} + \cdots\right)\left(\frac{\delta}{L}\right)^5 - \left(\frac{3026052}{67375}\mu - \frac{1277289}{1100}\frac{r}{L} + \cdots\right)\left(\frac{\delta}{L}\right)^6 - \left(\frac{5340897}{18865} + \cdots\right)\left(\frac{\delta}{L}\right)^7 + \cdots \quad (31)$$

where we assume that $\mu$ and $r/L$ are small and thus their higher powers were neglected. For very small deflection we by inverting Eq (31) obtain the following correction formula for calculation of bending stiffness

$$IE = \frac{FL^2}{48}\left(\frac{\delta}{L}\right)^{-1}\left\{1 - \left[\frac{12}{5}\mu + \frac{27}{2}\left(\frac{r}{L}\right)\right]\left(\frac{\delta}{L}\right) + \frac{72}{7}\left(\frac{\delta}{L}\right)^2 + \cdots\right\} \quad (32)$$

For numerical example we take $\mu = 0$, $r/L = 0.013$ and $\delta/L = 0.1$. In this case we obtain tangent angle $\phi_0 = 17^0$ which can be regard as small ( [10]) and bending stiffness *IE* which is about 8.5% greater than one calculated from elementary formula (1).

In Table 4 some comparison values for the case $r = 0$ and $\mu = 0$ are given. It is seen from the table that for relative error below say 3% the linear approximation is applicable for $\delta/L$ below say 0.06, the cubic and quantic approximation are applicable for $\delta/L$ below say 0.10 and polynomial approximation of seventh degree for $\delta/L$ below say 0.3. An impression how good various approximation are can be seen from Figure 12.

**Table 4.** Comparison between exact and approximate values of $FL^2/EI$ for given deflection $\delta/L$ calculated by polynomial approximation of various degree. Rerr is relative error in %. (* maximum load )

| $\delta/L$ | $\omega$ | exact | 1 | Rerr | 3 | Rerr | 5 | Rerr | 7 | Rerr |
|---|---|---|---|---|---|---|---|---|---|---|
| 0.02 | 0.34634 | 0.9561 | 0.960 | -0.41 | 0.956 | 0.00 | 0.956 | 0.00 | 0.956 | 0.00 |
| 0.05 | 0.54705 | 2.3393 | 2.400 | -2.59 | 2.338 | 0.04 | 2.339 | 0.02 | 2.339 | 0.00 |
| 0.10 | 0.77085 | 4.3377 | 4.000 | -10.66 | 4.306 | 0.72 | 4.320 | 0.40 | 4.338 | -0.01 |
| 0.15 | 0.93857 | 5.7567 | | | 5.534 | 3.87 | 5.641 | 2.00 | 5.762 | -0.10 |
| 0.20 | 1.07527 | 6.5119 | | | 5.650 | 13.23 | 6.712 | -3.07 | 6.538 | -0.40 |
| 0.24* | 1.16505 | 6.6718 | | | | | 7.305 | -9.49 | 6.714 | -0.63 |
| 0.30 | 1.29009 | 6.3340 | | | | | | | 6.161 | 2.73 |

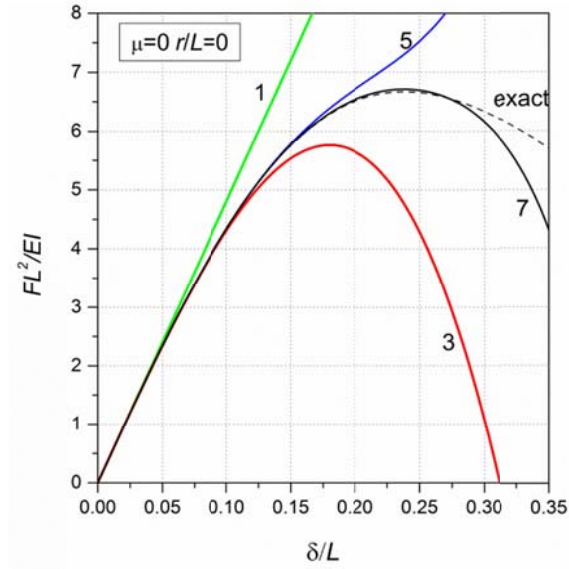

**Figure 11.** Comparison of load-deflection curve for various order of polynomial approximation.

At the end we note that in the similar way we can obtain also series for beam length and tangent angle. These approximations are

$$\frac{2\ell_0}{L} = 1 - 6\frac{r}{L}\frac{\delta}{L} + \frac{12}{5}\left(\frac{\delta}{L}\right)^2 - \left(\frac{24}{175}\mu - \frac{333}{35}\frac{r}{L}\right)\left(\frac{\delta}{L}\right)^3 - \frac{72}{35}\left(\frac{\delta}{L}\right)^4 + \cdots \quad (33)$$

$$\phi_0 = 3\left(\frac{\delta}{L}\right) - \frac{3}{10}\mu\left(\frac{\delta}{L}\right)^2 - \frac{27}{7}\left(1 - \frac{16}{225}\mu^2\right)\left(\frac{\delta}{L}\right)^3 + \frac{27}{56}\mu\left(1 - \frac{149}{225}\mu^2\right)\left(\frac{\delta}{L}\right)^4$$
$$+ \frac{24597}{2695}\left(1 - \frac{272}{4555}\mu^2\right)\left(\frac{\delta}{L}\right)^5 + \cdots \quad (34)$$

**Conclusion**

In the paper we present rather complete discussion of large deflections of beam under three-point bending. The main results are the following:

- The solution is expressed in terms of Jacobi elliptical functions which has theoretical and practical advantages over a solution expressed by elliptic integrals
- The domain of the solution is established. In particular it was shown that for $\sqrt{2}/2 \leq k < 0.9089$ or $0 \leq \mu < 0.86$ the load parameter is bounded to $0 \leq \omega \leq K(k)$.
- For $\sqrt{2}/2 \leq k < 0.9089$ the explicate formulas are given by which the required beam length and it's maximal deflection can be calculated (Eqs(22) and (23))
- Approximation expression for $FL^2/EI$ as polynomial function of $\delta/L$ were derived (Eq (31)).

At the end we note that the present solution is applicable for very flexible beams where effect of beam thickness can be neglected.

**References**

[1] M.A. Meyers, K.K. Chawla, W.F. Hosford, Mechanical behavior of materials, 2nd ed., Cambridge University Press, Cambridge, 2009.
[2] J.G. Freeman, Philosophical Magazine, Series 7, 37 (1946) 855-862.
[3] H.D. Conway, Phil. Mugazine, Series 7, 38 (1947) 905-911.
[4] R. Frisch-Fay, Flexible bars, Butterworths, London, 1962.
[5] D.C. West, Experimental mechanics, 21 (1964) 185-190.
[6] J.W. Westwater, P Am Soc Test Mater, 49 (1949) 1092-1118.
[7] P.S. Theocaris, S.A. Paipetis, S. Paolinelis, Journal of Testing and Evaluation, 5 (1977) 427-436.
[8] A. Ohtsuki, Bulletin of JSME, 29 (1986) 1988-1995.
[9] A.K. Arnautov, Mech Compos Mater, 41 (2005) 467-476.
[10] F. Mujika, Polym Test, 25 (2006) 214-220.
[11] M. Batista, International Journal of Solids and Structures, 51 (2014) 2308-2326.
[12] F.W.J. Olver, National Institute of Standards and Technology (U.S.), NIST handbook of mathematical functions, Cambridge University Press, Cambridge, 2010.
[13] B.C. Carlson, E.M. Notis, Acm Transactions on Mathematical Software, 7 (1981) 398-403.
[14] L.V. Ahlfors, Complex analysis : an introduction to the theory of analytic functions of one complex variable, 3d ed., McGraw-Hill, New York, 1979.